\documentclass[prd,amsmath,nofootinbib,amssymb,aps,twocolumn,preprintnumbers]{revtex4} 
\pdfoutput=1 

\usepackage{amsmath}
\usepackage{amssymb}
\usepackage{amsthm}
\usepackage{psfrag}
\usepackage{graphicx}
\usepackage{soul}

\newcommand\bw{\begin{widetext}}
\newcommand\ew{\end{widetext}}
\newcommand{\be}{\begin{equation}}
\newcommand{\ee}{\end{equation}}
\newcommand{\beqa}{\begin{eqnarray}}
\newcommand{\eeqa}{\end{eqnarray}}

\newcommand{\pd}{\partial}
\newcommand\m{\mu}
\newcommand\D{\Delta}
\newcommand\n{\nu}
\renewcommand\r{\rho}
\newcommand\s{\sigma}
\renewcommand\a{\alpha}
\renewcommand\b{\beta}
\renewcommand\l{\lambda}

\newcommand{\Gc}{G_{\text{cos}}}
\newcommand{\nn}{\nonumber}

\def\d{\partial}
\newcommand{\bseq}{\begin{subequations}}
\newcommand{\eseq}{\end{subequations}}

\newcommand{\di}{\mathrm d}

\begin{document}
\preprint{CERN-PH-TH/2014-196}
\preprint{INR-TH/2014-024}
\preprint{LAPTH-225/14}

\title{Cosmological constraints on deviations from
Lorentz invariance\\ in gravity and dark matter}
\author{B. Audren\,$^{a}$, D. Blas\,$^{b},$ M. M. Ivanov\,$^{c,d,f}$, J. Lesgourgues\,$^{a,b,e}$,
  S. Sibiryakov\,$^{a,b,d}$}
  \affiliation{{$^a$} \it FSB/ITP/LPPC,
 \'Ecole Polytechnique F\'ed\'erale de Lausanne,\\
 \normalsize\it CH-1015, Lausanne, Switzerland}
  \affiliation{{$^b$} \it Theory Group, Physics Department, CERN, CH-1211 Geneva 23,
 Switzerland}
  \affiliation{{$^c$} \it  Faculty of Physics, Moscow State University,  Vorobjevy Gory, 119991 Moscow, Russia}
    \affiliation{{$^d$} \it  Institute for Nuclear Research of the
Russian Academy of Sciences, 60th October Anniversary Prospect, 7a, 117312
Moscow, Russia}
\affiliation{{$^e$} \it LAPTH, U. de Savoie, CNRS,  BP 110, 74941 Annecy-Le-Vieux, France}
\affiliation{{$^f$} \it Sternberg Astronomical Institute, Universitetsky prospect, 13, 119992 Moscow, Russia}

\begin{abstract} 
We consider a scenario where local Lorentz invariance is violated by
the existence of a preferred time direction at every space-time point.
This scenario 
can arise in the context of quantum gravity and its description at low
energies contains a unit time-like vector field which parameterizes the
preferred direction.
The particle physics tests of Lorentz invariance preclude a direct
coupling of this vector to the fields of the Standard Model, but do not
bear implications for  dark matter. We discuss how
the presence of this  
vector and its possible coupling to dark matter affect the evolution 
of the Universe. 
At the level of homogeneous cosmology the only effect of 
Lorentz
invariance violation 
is a rescaling of the expansion rate.
The physics is richer at the level of
perturbations. 
We identify three effects
crucial for observations: the rescaling of the matter contribution to
the Poisson equation, the appearance of an extra contribution to the
anisotropic stress and  
the scale-dependent enhancement of dark matter clustering. 
These effects result in distinctive 
features in the power spectra of the CMB and 
density fluctuations. 
Making use of the data from Planck and
WiggleZ we obtain 
the most stringent cosmological constraints to date 
on departures from 
Lorentz symmetry.
Our analysis provides the first direct bounds on
deviations from Lorentz invariance in the dark matter sector.
\end{abstract}

\maketitle

\section{Introduction}
Understanding the properties of the different
components  of the Universe  
is one of the major aims of cosmology
and particle physics. All  current data are compatible
with the $\Lambda$CDM scenario, where the fields of the Standard Model
of particle physics (SM) are supplemented by a dark matter component
and a cosmological constant. 
This scenario assumes that Lorentz invariance (LI) is a fundamental
property of  
Nature, and thus holds for all sectors of the theory.
As for any postulate, it is important to verify this assumption
experimentally. Indeed, many tests have been performed within SM
establishing the validity of LI in this sector with exquisite accuracy
\cite{Kostelecky:2008ts,Liberati:2013xla}. On the other hand, the
tests of LI in
gravity are mostly limited to the framework of the post-Newtonian 
expansion and are based on the observations in a rather narrow range
of distances relevant for the planetary and stellar dynamics
\cite{Will:2014kxa}. The probes going beyond the post-Newtonian
description, such as the study of energy loss rate by binary pulsars,
are able to give only mild bounds on deviations from LI in gravity
\cite{Yagi:2013qpa}. Moreover, we do not have any direct
information about the validity of LI in the dark sectors of the
Universe: dark matter and dark energy. Given the key role played by LI
in describing Nature it is essential to test it at all accessible
scales for as many observables as possible.  

Further motivation to consider Lorentz invariance violation (LV)
comes from the quest 
for a theory of quantum gravity.  
A concrete scenario in this direction is Ho\v rava gravity
\cite{Horava:2009uw,Blas:2009qj}, 
where the ultraviolet behavior of the gravitons
is modified by operators  
breaking LI; this leads to a power counting renormalizable theory. In
a 
broader perspective, one could envisage that the theory of
quantum gravity would generate LV at low-energies, which  
one could parameterize using effective theories, such 
as the Einstein-aether model 
\cite{Jacobson:2000xp} or the SM extension of particle physics
\cite{Colladay:1998fq}. 
These theories generically include  extra massless excitations in the
gravity sector which yield a very rich phenomenology beyond 
general relativity (GR) at all scales, in particular, in cosmology.
Given the remarkable progress in the amount and quality of the
cosmological data during recent years, it is reasonable to ask if they
can provide information about the validity of LI in gravity and the
dark sectors. This topic was first addressed in \cite{Zuntz:2008zz} where LV
in gravity parameterized by the Einstein-aether model was
constrained using the data on cosmic microwave background (CMB) and
large scale structure (LSS). More recently \cite{Audren:2013dwa}
considered a scenario with LV in gravity and dark energy
sectors, which presents an alternative to $\Lambda$CDM where cosmic
acceleration is sourced by a term  
insensitive to large ultraviolet corrections \cite{Blas:2011en}. 

The purpose of this paper is to use the recent results
on the CMB \cite{Planck} and 
linear matter power spectrum (LPS) \cite{WiggleZ} to constrain the
deviations with respect to $\Lambda$CDM associated with LV in gravity
and {\em dark matter}. 
We will focus on the effects from the existence of a preferred
time-direction. Our study is based 
on the phenomenological description of LV in
dark matter developed in \cite{Blas:2012vn}.

The paper is organized as follows. In Sec.~\ref{sec:2} we summarize
the formalism for the effective description of LV in gravity and dark
matter. We discuss the cosmological background evolution and 
present the equations for the perturbations in Sec.~\ref{sec:cosmo}.
In Sec.~\ref{sec:effects} we describe the effects of LV on the CMB
and LPS.
Sec.~\ref{sec:5} contains the main results of the paper: the
observational constraints on the LV parameters. 
We conclude in Sec.~\ref{sec:6}.

\section{Lorentz breaking theories of gravity and dark matter}
\label{sec:2}
We assume that the
Universe is permeated by a time-like vector field $u_\m$ ---
``aether'' --- defining a
preferred time direction at every point of space-time. This vector is
taken to have unit norm\footnote{We use the $(-+++)$
signature for the metric and work in units $\hbar=c=1$.},
\be
\label{unnorm}
u_\m u^\m=-1.
\ee
The presence of this field breaks LI locally down to spatial rotations.
The covariant action for $u_\m$ and the metric $g_{\m\n}$ with the
minimum number of derivatives reads,
\be
\begin{split}
\label{aetheract}
S_{[{\rm EHu}]}=\frac{1}{16\pi G_0}\int \di^4x \sqrt{-g}
\big[R-&K^{\m\n}_{\phantom{\m\n}\s\r}\nabla_\m u^\s\nabla_\nu u^\rho\\
&\quad+l (u_\m u^\m+1)\big]\,,
\end{split}
\ee
where $R$ is the Ricci scalar
for the metric $g_{\m\n}$, 
\be
\label{Kmnsr}
K^{\m\n}_{~~~\s\r}\equiv c_1g^{\m\n}g_{\s\r}+c_2\delta_\s^\m\delta_\r^\n
+c_3\delta_\r^\m\delta_\s^\n-c_4u^\m u^\n g_{\s\r},
\ee
and $l$ is a Lagrange multiplier that enforces the unit-norm
constraint. This is the action of the Einstein-aether model
\cite{Jacobson:2000xp,Jacobson:2008aj}. 
The parameter $G_0$ in (\ref{aetheract}) 
is related to Newton's constant as \cite{Jacobson:2008aj,Blas:2009qj}
\be
\label{GN}
G_N\equiv G_0\big(1-(c_4+c_1)/2\big)^{-1}.
\ee 
The dimensionless constants $c_a$,
$a=1,2,3,4$, 
characterize the strength of the interaction of the aether  $u_\m$
with gravity.  

One can require additionally that the field $u_\m$ is orthogonal to a
family of three-dimensional hypersurfaces defined as the leaves of
constant scalar field  
$\varphi$, 
\be
u_\m\equiv\frac{\pd_\m \varphi}{\sqrt{-\pd^\n \varphi \pd_\n \varphi}}\;.
\ee
Then  the action (\ref{aetheract}) corresponds to the 
{\it khronometric} model which represents the
low-energy limit
of Ho\v rava gravity \cite{Blas:2009qj,Blas:2010hb}. In this case the
term with the Lagrange multiplier is redundant and the four terms in
(\ref{Kmnsr}) are linearly dependent. Thus 
 $K^{\m\n}_{\phantom{\m\n}\s\r}$ can be reduced to its last three terms with coefficients
\be
\label{khpar}
\l\equiv c_2,~~~\beta\equiv c_3+c_1,~~~\alpha\equiv c_4+c_1\;.
\ee
The detailed relation between the Einstein-aether and khronometric
models has been worked out in \cite{Jacobson:2013xta}.

The previous action should be supplemented by an action for the matter
sector. 
This consists of the SM part and a dark matter (DM)
component. Generically, both can be directly coupled to the vector
$u_\m$. 
For the SM part, this coupling is strongly constrained
from tests of LI in particle physics experiments 
\cite{Kostelecky:2008ts,Liberati:2013xla} 
and checks of the weak equivalence principle \cite{Will:2014kxa}, 
which implies that it is negligible in cosmology. 
In what follows we
assume that there is no direct coupling between SM and the
aether.
This decoupling can be ensured, e.g., by imposing (softly broken) supersymmetry  \cite{GrootNibbelink:2004za,Pujolas:2011sk}  
or by a dynamical emergence of LI at low energies in the SM sector \cite{Bednik:2013nxa}.
In contrast, these mechanisms do not work for particles that are singlets under the gauge group (in the case of supersymmetry) or for weakly coupled particles (in the case of dynamical emergence). 
As DM must be weakly interacting with SM and is conventionally believed to be a gauge singlet, its coupling to the aether can be sizeable.

Since the relation between dark DM and SM particles is not established, the only direct tests of LV in DM can come from cosmological analysis.
For DM, the possibility of a direct coupling to $u_\m$ was first
considered in \cite{Blas:2012vn}. It was shown that within the
fluid description of DM this coupling leads to the following effective
action,
\be
\label{SDMu}
S_{[{\rm DMu}]}=-m\int \di^4 x \sqrt{-g}\, n\, F(u_\mu v^\mu), 
\ee
where $m$ is the mass of the DM particles\footnote{As explained in
  \cite{Blas:2012vn}, this framework can be generalized almost without
changes to the case of any DM admitting a fluid description, such
as, for example, axionic DM. We stick in this paper to
the simple physical picture of weakly interacting massive particles
for concreteness.}, 
$n$ is their number density and
$v^{\mu}$ is their four-velocity. The function   
$F(u_\mu v^\mu)$ parameterizes the interaction between the DM fluid
and the aether; without loss of generality, it can be normalized to $F(1)=1$. 
We will see below that the effect on cosmology is encapsulated by a
single parameter,
\be
\label{YY}
Y\equiv F'(1)\;.
\ee

In particle physics LV is usually associated with the modification of
the particles' dispersion relations ---
the dependence of their energy on the spatial momentum. 
The effective field theory framework predicts that 
at relatively low energies the leading modification occurs in the
quadratic term 
\cite{Coleman:1998ti},\
giving
\be
\label{disprel}
E^2=m^2+(1+\xi) {\bf p}^2\;,
\ee
where for a relativistic theory $\xi=0$.  
Requiring that the DM particles have
the dispersion relation (\ref{disprel}) in the rest frame of the
aether (i.e. where it has the form $u_\m=(1,0,0,0)$) corresponds to
choosing a
function $F$ in (\ref{SDMu}) with 
$Y=\xi/(1+\xi)$ \cite{Blas:2012vn}. Thus, by
putting bounds on the parameter $Y$ we will be able to constrain the
deviations of the DM dispersion relation from the relativistic form.

In deriving the
equations of motion following from (\ref{SDMu}) the variation of the fields 
must be subject to the constraints\footnote{Alternatively, within the
  so-called pull-back formalism, one introduces a triple of scalar
  fields parameterizing the fluid elements and varies with respect to
  these fields without any restrictions
  \cite{Andersson:2006nr,Dubovsky:2005xd,Blas:2012vn}.} $v^\m v_\m=-1$, 
$\nabla_\m(n\,v^\m)=0$, the latter expressing the particle number
conservation. The resulting equations for the aether-DM system can be
found in \cite{Blas:2012vn}. 

Finally, we add to the total energy budget of the Universe the
cosmological constant $\Lambda$ as the source of the cosmic
acceleration. We will refer to the resulting cosmological model as
$\Lambda$LVDM. 

Various combinations of the parameters introduced above are
constrained by experiment. Despite the fact that the aether does not
couple directly to SM, it affects the gravitational interactions among
celestial bodies. 
The Solar System tests provide the 
bounds $|\alpha_1|\lesssim 10^{-4}$ and $|\alpha_2|\lesssim 10^{-7}$,
where $\alpha_1$, $\alpha_2$ are certain combinations of the parameters
$c_a$ entering in the post-Newtonian dynamics 
\cite{Jacobson:2008aj,Blas:2010hb,Blas:2011zd}.
Since these bounds are much stronger than those
expected from cosmology, we will impose vanishing of $\alpha_1$,
$\alpha_2$ as priors in our parameter extraction
procedure. 
 In terms of the original coefficients this amounts to
imposing the following relations\footnote{The difference between the
  Einstein-aether and khronometric models is due to the helicity-1
  perturbations in the Einstein-aether case, which contribute 
  into $\a_1$, $\a_2$.}:\\
~\\
- for Einstein-aether model:
\bseq
\begin{align}
c_1c_4+c_3^2=0\;,\\
\label{consaeth}
2c_1+3c_2+c_3+c_4=0\;;
\end{align}
\eseq
- for khronometric model:
\be
\label{conskhron}
\a-2\b=0\;.
\ee
Remarkably, in the khronometric case a single relation
(\ref{conskhron}) suffices to ensure vanishing of all post-Newtonian (PN)
parameters. 

A stringent bound $|\hat\alpha_2|\lesssim 10^{-9}$ has been
derived from the dynamics of solitary pulsars \cite{Shao:2013wga},
where $\hat\alpha_2$ is the generalization of $\alpha_2$ for strong
gravitational fields. However, this translates in rather mild
constraints on the parameters of the model once the vanishing of the PN
parameters, eqs. (\ref{consaeth}) or (\ref{conskhron}), is imposed 
\cite{Yagi:2013qpa}.
Stronger constraints follow from the radiation
damping in binary systems \cite{Yagi:2013qpa}, which is 
an efficient way to 
test LV beyond the combinations $\alpha_1$ and $\alpha_2$. 
Except for some degeneracies, these bounds are of order $|c_a|\lesssim
10^{-2}$.
Besides, there are theoretical constraints that further restrict the
allowed parameter space, see e.g. \cite{Yagi:2013qpa}
for a succinct review. In particular, stability imposes the following
positivity conditions on the combinations (\ref{khpar}),
\be
\label{posconds}
0<\a<2~,~~~~0 < \b+\l \;.
\ee  

A future direct detection of DM, implying its appreciable coupling to
 the SM fields, can lead to strong constraints on the DM -- aether
 interaction. In this case, following the arguments of
 \cite{Bovy:2008gh,Carroll:2008ub} one 
 would be able to estimate the direct coupling of SM to $u_\m$ induced
 by radiative corrections due to DM loops and apply the stringent
 constraints on LV in SM. However, these bounds will be
 model-dependent and, in the absence of a direct DM detection so far,
 we do not take them into account in the present work.

\section{Cosmological background and perturbations }
\label{sec:cosmo}
It was shown in
\cite{Blas:2012vn} that the only effect  
of the previous modifications for homogeneous and isotropic
backgrounds is a rescaling of the gravitational
constant in the Friedmann 
equation\footnote{In particular, the introduction of the aether and its 
interaction with DM cannot, by itself, provide the present accelerated expansion of the universe.  
More ingredients can be added to this setup to 
realize dark energy in a technically natural way \cite{Audren:2013dwa} .}
\be
G_{cos}\equiv G_0\left[1+\frac{c_1+3c_2+c_3}{2}\right]^{-1}.
\ee  
This does not coincide with the Newton's constant governing local gravity \eqref{GN}. Since $G_{cos}$ affects the relic abundance of different elements in the Universe, 
it can be constrained with the Big Bang Nucleosynthesis (BBN) 
data \cite{Carroll:2004ai}. Note, however, that in the Einstein-aether
case the condition for vanishing of the post-Newtonian parameters
(\ref{consaeth}) implies that $G_{cos}$ coincides with $G_N$ to the
linear order in the aether parameters $c_a$ and the BBN constraints
are weakened.

For the perturbations, we will focus on the scalar sector of the
theory\footnote{Vectors may be important in the
  Einstein-aether case \cite{ArmendarizPicon:2010rs,Nakashima:2011fu} 
if they are efficiently produced in the primordial Universe and do not
decay with time. This happens in a restricted portion of the parameter
space.}. 
The dynamic of this sector is identical in the Einstein-aether and
khronometric cases and depends only on the combinations
(\ref{khpar}). We work in the synchronous gauge \cite{Ma:1995ey} where
the metric takes the form,
\be
\label{synchg}
ds^2\!=\!a^2(\tau)\!\bigg[\!-d\tau^2+\bigg(\!\delta_{ij}
+\frac{\d_i\d_j}{\D}h
+6\bigg(\!\frac{\d_i\d_j}{\D}-\frac{\delta_{ij}}{3}\!\bigg)\eta
\bigg)dx^i dx^j\!\bigg],
\ee
where $\D\equiv\d_i\d_i$ is the spatial Laplacian.
Given eq.~(\ref{unnorm}), the scalar perturbations of the field $u_\m$ can be parametrised as
\be
u_0=a(\tau), \ u_i=a(\tau)\partial_i\chi\;.
\ee
In Fourier space, the DM equations read
\bseq 
\label{pertdm}
\begin{align}
\label{pertdm1}
\dot{\delta}_{[dm]}+\theta_{[dm]}+\frac{\dot{h}}{2}=0\;,\\
\label{pertdm2}
\dot\theta_{[dm]}+{\cal H}\theta_{[dm]}+\frac{Yk^2}{1-Y}(\dot\chi+
{\cal H}\chi)=0\;,
\end{align} 
\eseq
where dot stands for the derivatives with respect to the conformal
time $\tau$; ${\cal H}\equiv\dot a/a$; the DM density contrast and
velocity divergence are defined in the usual way,
\be
\delta_{[dm]}\equiv\frac{\delta\rho_{[dm]}}{\rho_{[dm]}}\,,
\qquad
\theta_{[dm]}\equiv \frac{ik_jv^j_{[dm]}}{a(\tau)}\,,
\ee 
and 
$Y$ has been defined in (\ref{YY}).
Unlike the standard case we cannot put the DM velocity to zero by a
residual gauge choice. The reason is that, due to the interaction with
the aether, DM does not follow geodesics. Instead, from
(\ref{pertdm2}) we see that we can impose
\be
\label{gauge}
\theta_{[dm]}=-\frac{Yk^2}{1-Y}\chi\,.
\ee
In this gauge the equations for
$u_\mu$ reduce to
\be 
\label{eq:chi}
\begin{split}
  \ddot{\chi}& =  -\frac{c_\chi^2}{2}\dot h - 2\frac{\beta}{\alpha}\dot \eta- 2 {\cal H} \dot{\chi} \\
   &-\left[(1+B) {\cal H}^2 + (1-B) \dot{\cal H}+c_\chi^2k^2
+\frac{c_{\chi}^2 k_{Y,0}^2}{a}  \right] \chi\,,
\end{split}
\ee
with $B\equiv\frac{\beta+3\lambda}{\alpha}$, 
\be
\label{ckY}
c_\chi^2\equiv \frac{\beta+\lambda}{\alpha}\,,\qquad
k^2_{Y,0}\equiv\frac{3Y\Omega_{dm}H_0^2}{(\b+\l)(1-Y)}\frac{G_0}{\Gc}\;,
\ee
where $\Omega_{dm}$ is the dark matter fraction and $H_0$ is the
Hubble parameter today.
The constant $c_\chi^2$ has the physical meaning of the squared
velocity of the longitudinal aether waves, its positivity is guaranteed by
the conditions (\ref{posconds}). From theoretical viewpoint, 
$c_\chi$ can be both
smaller or larger than unity: superluminal propagation is compatible
with causality in the presence of LV. Importantly, the absence of energy
losses by ultra-high energy cosmic rays via vacuum Cherenkov emission
of the
$\chi$-field requires that $c_\chi$ must be equal or bigger 
than
1~\cite{Elliott:2005va}.\footnote{More precisely, the bound reads 
$c_\chi^2>1-10^{-22}$ as long as the process would occur with sizable
probability over the cosmological distances.  This is the case as long as 
$\frac{(\b-\a)^2}{\a}>10^{-30}$.} We
will impose this requirement as a prior in our parameter extraction
procedure. 

The last term in the square brackets in (\ref{eq:chi}) effectively
introduces a (time-dependent) mass for the aether perturbations.
We impose a theoretical prior that the square of this mass must be
positive, $k_{Y,0}^2>0$ --- otherwise one expects rapid instabilities
in the aether--DM sector \cite{Blas:2012vn} in contradiction with
observations. Thus in what follows we restrict to 
\be
\label{Yprior}
0\leq Y< 1\;.
\ee 
On the other hand, the positive effective mass 
leads to the suppression of the aether perturbations 
for $k<k_{Y,0}/\sqrt{a(\tau)}$ implying that LV effects are screened
at distances longer than
$2\pi k_{Y,0}^{-1}$ \cite{Blas:2012vn}.\footnote{Note that the definition of
  $k_{Y,0}$ used in this paper differs from that in \cite{Blas:2012vn}
by a factor of $c_\chi^{-2}$.} 

Finally, the only two independent equations following from the
linearized Einstein equations are,
\begin{align}
k^2\eta - \frac{1}{2}\frac{G_0}{\Gc}\mathcal H \dot h = &-4\pi a^2 G_0
\sum_{i}  \rho_i\delta_i 
 \nn\\
    &- {\alpha k^2} (\mathcal H (1-B)\chi + \dot\chi ), 
  \label{eq:T00}\\
  \label{eq:Tii}
  \ddot h = -2\mathcal H \dot h + \frac{\Gc}{G_0}& 2k^2 \eta-24\pi a^2 \sum_{i}\delta  p_i\nn\\
   -\alpha B  \frac{\Gc}{G_0} k^2(\dot \chi &+ 2\mathcal H\chi)\,,
\end{align}
where the sums on the r.h.s. run over the matter species filling the
Universe: DM, baryons and radiation (including neutrinos, which we
assume to be massless). The system is completed by the
standard equations for the baryon and radiation components
\cite{Ma:1995ey}.  

To solve the previous equations for a given Fourier mode $k$ we set
the initial conditions in the radiation era, at a moment $\tau_0$ when
the wavelength associated to $k$ is well outside the Hubble scale,
i.e. when $k\ll {\cal H}$.  These modes will be initiated in the adiabatic
growing mode of the model \cite{Blas:2012vn}, which under broad
assumptions gives the dominant contribution to 
the 
perturbations\footnote{Isocurvature modes were considered in \cite{ArmendarizPicon:2010rs,Blas:2011en,Blas:2012vn}. 
In the khronometric case they always decay outside the horizon once the  
PPN condition \eqref{conskhron} is imposed.
In the Einstein-aether model they also decay if $Y\neq 0$ and the LV parameters satisfy \eqref{consaeth}.
Finally, if 
$Y=0$ the isocurvature modes stay constant at super-horizon distances in the Einstein-aether case after
imposing \eqref{consaeth}. Even a small deviation
from \eqref{consaeth} means that the modes either grow or decay \cite{Blas:2011en}. Thus, if primordially generated,  isocurvature modes can survive only in the Einstein-aether theory with LI DM \cite{ArmendarizPicon:2010rs}. We leave the study of their phenomenology for future.
It is worth mentioning that the presence of the aether does not necessarily lead to the production of isocurvature modes during inflation. There are examples of inflationary models (see \cite{Ivanov:2014yla}) containing the aether and generating primordial perturbations in pure adiabatic modes. 
}. 
Further details on the numerical procedure 
in a similar theory can be found in
\cite{Audren:2013dwa}.

\section{Effects on observations}
\label{sec:effects}

Some of the cosmological effects of LV were
discussed in the past
\cite{Carroll:2004ai,Zuntz:2008zz,Kobayashi:2010eh,ArmendarizPicon:2010rs,
Blas:2012vn,Audren:2013dwa}. 
Here 
we summarize and extend them emphasizing the effects related to LV in
dark matter. 

The deviations of $\Lambda$LVDM from the standard cosmology can be
divided into three categories:
\begin{itemize}
\item[{\em (i)}] effects related to the difference between $G_N$ and
  $G_{cos}$; they are proportional to 
\be 
\label{selfgravity}
\frac{G_N}{G_{cos}}-1=\frac{\a+\b+3\l }{2}+ \mathcal{O}(\a^2)\;,
\ee
where by $O(\a^2)$ we mean any subleading contributions in the 
parameters $\a,\b,\l$
\item[{\em (ii)}] effects due to the presence of shear; they are proportional to $\b$
  \cite{Blas:2012vn} 
\item[{\em (iii)}] effects due to LV in DM appearing for non-zero value of
  the parameter $Y$.
\end{itemize}
The first two classes of effects are common to a very broad class of
modified gravity models, not necessary based on the Einstein-aether
\cite{Clifton:2011jh,Amendola:2012ys,Saltas:2014dha}. A detailed study
of their impact on the CMB and matter power spectrum was performed in
\cite{Audren:2013dwa} with the following outcome:

{\em (i)} Whenever $G_{cos}\neq G_N$  
the Poisson equation for sub-horizon scales in an expanding background
is modified. As a result, the growth of sub-horizon perturbations is
uniformly enhanced. For instance, during  matter domination the density
contrast behaves as (cf. \cite{Kobayashi:2010eh})
\[
\delta \propto a^{\frac{1}{4}(-1+\sqrt{1+24 G_N/G_{cos}})}\,.
\]
This increases the amplitude and changes the slope
of the LPS. As for the CMB, the main changes are a shift of the acoustic
peaks due to an increase of the gravitational potential of the
primordial plasma, and an enhancement of the integrated Sachs--Wolfe
(ISW) effect at intermediate multipoles $10\lesssim l \lesssim
100$. Note that the effects of enhanced gravity vanish to the leading
order in the LV parameters in the Einstein-aether model once the
condition (\ref{consaeth}) of the absence of PN corrections is
imposed. Indeed, in terms of the parameters $\a,\b,\l$ the latter
condition takes the form,
\be
\label{consaeth1}
\a+\b+3\l=0\;,
\ee 
which is exactly the combination appearing in (\ref{selfgravity}).
On the other hand, in the khronometric case, enhanced gravity is
compatible with the PN constraint (\ref{conskhron}).

{\em (ii)} The mode $\chi$ produces shear at superhorizon scales,
which decays at later times. 
Interestingly, a non-zero
coupling between the aether and DM postpones this decay \cite{Blas:2012vn}.
The shear
smoothes out the metric perturbations
and leads to an overall suppression of the
CMB anisotropies and LPS at the scales corresponding to the
sound speed of the $\chi$-mode, $c_{\chi}$. This effect is partially degenerate with an
overall rescaling of the amplitude of the primordial fluctuations and
is only weakly constrained by the data. 

\begin{figure}[t]
\begin{center}
\includegraphics[width=0.44\textwidth]{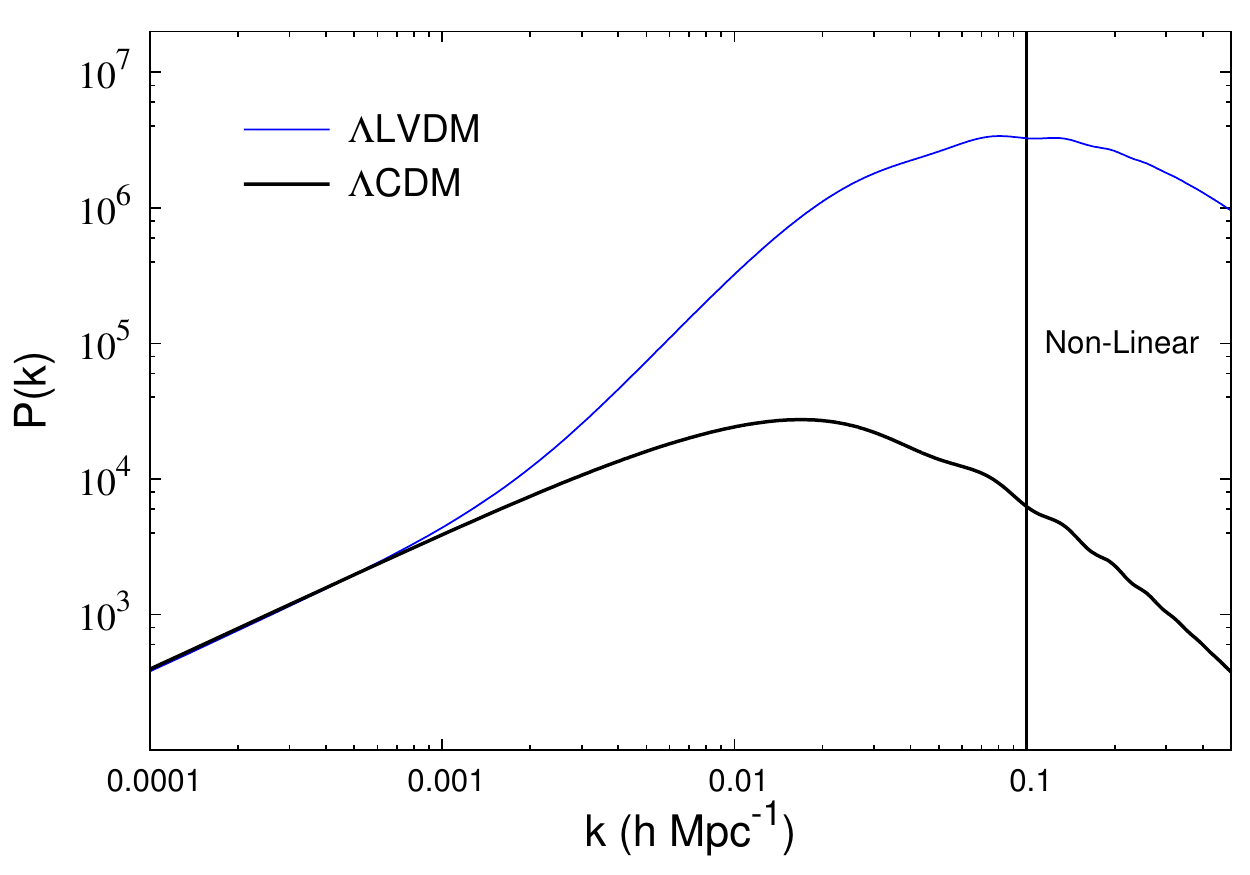}
\includegraphics[width=0.45\textwidth]{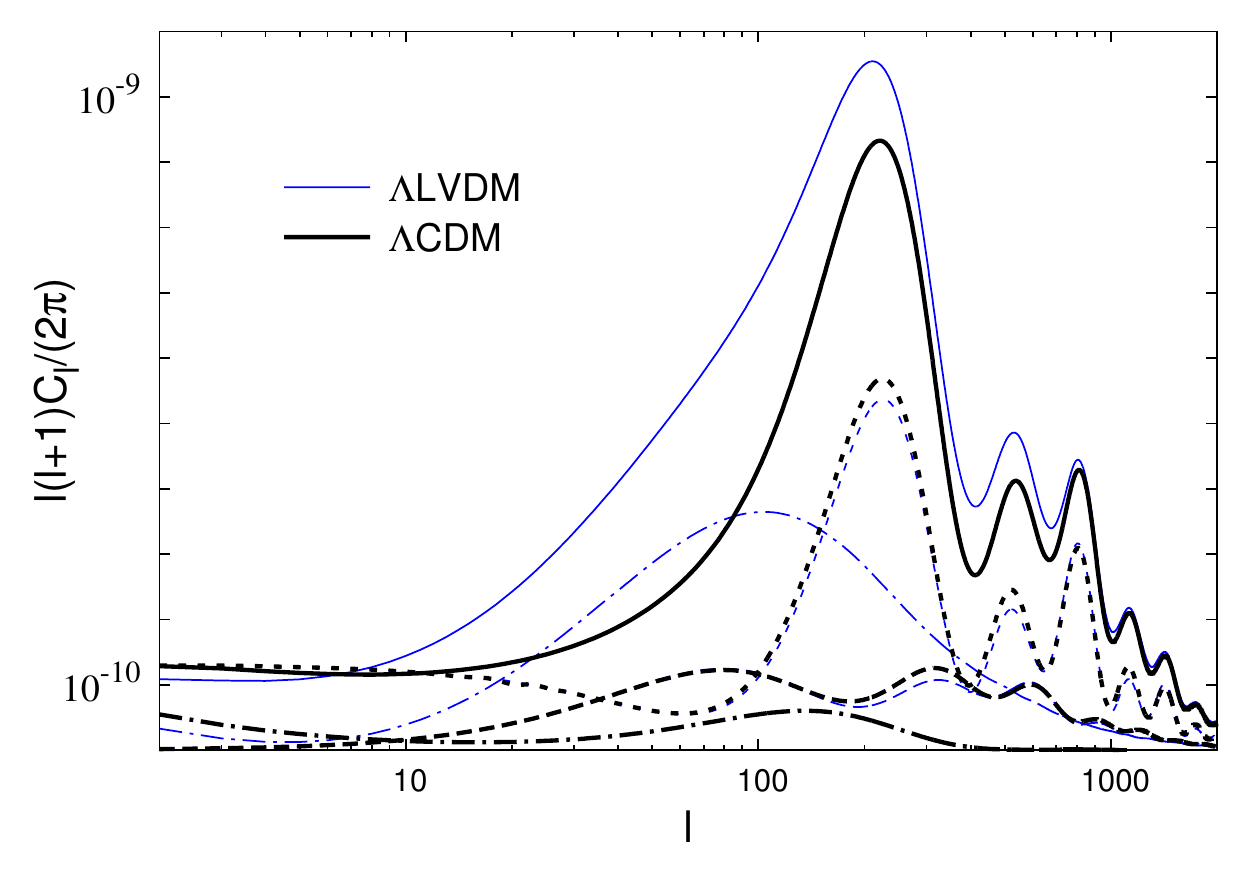}
\end{center}
\caption{\label{fig:deg} 
\textit{Upper panel:} Linear matter power spectrum in $\Lambda$CDM (thick
black line) and $\Lambda$LVDM models (thin blue line) at redshift $z=0$. \\ 
\textit{Lower panel:} Temperature anisotropy spectrum (solid) and its 
decomposition in terms of Sachs--Wolfe (dotted), Doppler (dashed) and
integrated Sachs--Wolfe (dot-dashed) contributions. Thick black lines
represent the $\Lambda$CDM model, while thin blue lines are used for
the $\Lambda$LVDM model. 
}
\end{figure}

Concerning the effects of type 
{\em (iii)}, the main consequence of LV in DM is a scale-dependent
modification of the growth rate of the density perturbations
\cite{Blas:2012vn}. For modes with $k>k_{Y,0}/\sqrt{a(\tau)}$ the
coupling to the aether violates the equivalence principle in the DM
sector: the inertial mass of DM particles is smaller than their
gravitational mass. This leads to an enhanced growth of the density
contrast during matter domination,   
\be
\label{unscreen}
\delta_{[dm]} \propto
a^{\frac{1}{4}\big(-1+\sqrt{25+\frac{24\Omega_{dm}Y}
{(\Omega_{dm}+\Omega_{b})(1-Y)}}\big)}
\,. 
\ee
On the other hand, the violation of the equivalence principle in DM is
screened for modes 
with $k < k_{Y,0}/\sqrt{a(\tau)}$ and is almost absent at radiation
domination. Thus, the total effect is a change of the slope of the LPS at 
$k>k_{Y,0}$. This is clearly visible in Fig.~\ref{fig:deg}, upper
panel, where we display the results of a numerical 
simulation of the LPS
in the $\Lambda$LVDM model using the modified Boltzmann code 
{\sc Class} \cite{Blas:2011rf}.
The numerical values of the model parameters\footnote{
For the standard cosmological parameters we take:
$n_s=1,\; h=0.7,\;\Omega_b=0.05,\; \Omega_{dm}=0.25,\;
A_s=2.3\cdot10^{-9}\,,$ and neglect the effects of 
reionization for 
illustrative purposes when computing the CMB spectrum.}  were 
chosen in a way to switch off the effects of enhanced gravity {\em (i)} and minimize 
the effects of the shear~{\em (ii)},
\be
\label{refpar}
\a=0.005,~~\b=0.025,~~ \l=-0.01,~~ Y=0.5 \,.
\ee
This corresponds to $k_{Y,0}=1.14 \cdot 10^{-3}\, h\,\text{Mpc}^{-1}$.

A comment on the definition of the power spectrum is in order. This
definition is not obvious in the $\Lambda$LVDM model where both gravity and
the dynamics of DM are modified leading to non-trivial relations
between the density contrasts of DM, baryons and the fluctuations of
the gravitational potential. In particular, the violation of the
equivalence principle by DM at scales $k>k_{Y,0}/\sqrt{a}$ is
accompanied by a bias factor \cite{Blas:2012vn},
\be
\label{bias}
\frac{\delta_{[b]}}{\delta_{[dm]}}=1-Y\,.
\ee
Therefore taking $\delta_{[dm]}$ or $\delta_{[b]}$ to compute the power
spectrum would produce different results. Following
\cite{Audren:2013dwa} we use the Poisson equation to {\em define} the
total density perturbation,
\be
\label{Poisson}
\delta\rho_{[tot]}=-\frac{k^2\phi}{4\pi G_N a^2}\;,
\ee
where $\phi$ is the perturbation of the Newtonian potential related to
$\eta$ and $h$ by the standard formulas \cite{Ma:1995ey}. Then the
power spectrum is defined as the ratio of this density perturbation to
the {\em apparent} background density of DM and baryons. The latter
differs from the true density entering the Friedmann equation due to
the difference between $G_{cos}$ and $G_N$. Taking this into account,
one obtains the final formula for the power spectrum \cite{Audren:2013dwa},
\be
\label{powspec}
P(k)=\bigg(\frac{G_N}{G_{cos}}\bigg)^2
\frac{\langle|\delta\rho_{[tot]}(\vec{k})|^2\rangle}
{\rho_{[dm]}+\rho_{[b]}}\;.
\ee 
The definition (\ref{Poisson}) has a limited applicability: it is
meaningful only well inside the Hubble horizon and for negligible
shear. However, both conditions are satisfied for the range of
wavenumbers where the power spectrum is actually measured. Taking the
gravitational potential as the basis for the definition of LPS is
motivated by the fact that this quantity is directly probed by the
lensing surveys. The amount of galaxies and clusters is also believed
to trace linearly the underlying gravitational field at large scales 
($k<0.1\,h\,\text{Mpc}^{-1}$), though contamination by a scale-dependent
bias cannot be excluded (cf. \cite{Villaescusa-Navarro:2013pva}). A
dedicated study of the clustering dynamics in the model at hand is
required to definitely resolve this issue, which is beyond the scope
of this paper.

A cleaner observable, insensitive to the above ambiguities, is the
spectrum of CMB anisotropies. The temperature spectrum computed
using {\sc  Class} in the model with the parameters (\ref{refpar}) is shown in
the lower panel of Fig.~\ref{fig:deg}. 
The main modification comes from
the very strong ISW effect induced by the 
accelerated growth of perturbations \eqref{unscreen}.
It extends over a wide 
range of multipoles from $l\sim k_{Y,0}\tau_0$ ($\approx 25$ for 
the chosen parameters) up to $l\sim 1000$. 
Modifications in the Sachs--Wolfe and Doppler effects are small,
consistent 
with the decoupling of DM
from photon-baryon plasma at the epoch of recombination
\cite{Voruz:2013vqa}.
In particular, the positions of the acoustic peaks are not changed, which
distinguishes the $\Lambda$LVDM impact on CMB from that of the
modified Poisson equation {\em (i)}
helping to break the degeneracy between them.

\section{Comparison with data}
\label{sec:5}
\begin{table*}[t]
  \begin{tabular}{|lccccccccc|}
     \hline
     &$100~\omega_{b }$  & $\omega_{cdm }$  & $n_{s }$  &
     $10^{+9}A_{s }$  & $h$  & $z_{reio }$  & $\alpha$  &
     $c_{\chi}^2$  & $Y$ \\ \hline 
     E-ae \; & $2.225_{-0.031}^{+0.028}$ & $0.1178_{-0.0022}^{+0.0025}$ &
     $0.9635_{-0.0073}^{+0.0068}$ & $2.159_{-0.050}^{+0.048}$ &
     $0.683^{+0.010}_{-0.012}$ & $10.4_{-1.0}^{+1.1}$ &
     $<5.0 \cdot 10^{-3}$ & $<240$ & $<0.028$$$ \\[0.2cm]
         kh \; & $2.227_{-0.031}^{+0.028}$ & $0.1174_{-0.0022}^{+0.0025}$ &
     $0.9639_{-0.0072}^{+0.0068}$ & $2.152_{-0.054}^{+0.048}$ &
     $0.685^{+0.010}_{-0.012}$ & $10.4_{-1.0}^{+1.1}$ &
     $<1.1 \cdot 10^{-3}$ & $<55$ & $<0.029$$$ \\[0.2cm]
       E-ae, $Y\equiv 0$ \; & $2.207_{-0.027}^{+0.026}$ &
       $0.1200_{-0.0019}^{+0.0019}$ & 
     $0.9598_{-0.0063}^{+0.0062}$ & $2.182_{-0.051}^{+0.045}$  &
     $0.673^{+0.009}_{-0.009}$ & $10.1_{-1.0}^{+1.0}$ &
     $<1.0 \cdot 10^{-2}$ & $<427$ & $-$$$ \\[0.2cm]
       kh, $Y\equiv 0$  \; & $2.212_{-0.028}^{+0.027}$ & $0.1191_{-0.0020}^{+0.0021}$ &
     $0.9607_{-0.0066}^{+0.0067}$ & $2.161_{-0.054}^{+0.053}$ &
       $0.677^{+0.009}_{-0.011}$ & $10.6_{-1.0}^{+1.1}$ &
     $<1.8 \cdot 10^{-3}$ & $<91$ & $-$$$ \\[0.2cm]
     \hline
   \end{tabular}
\caption{Mean values and 68\% CL minimum credible
interval for the parameters of the $\Lambda$LVDM models. The first and
third lines correspond to the Einstein-aether and
the second and fourth lines --- to the
khronometric cases. For $\a$, $c_\chi^2$ and
$Y$ we give 95\% CL upper limits. The bounds in the first two lines are subject to the priors (\ref{priors}).}
\label{table}
\end{table*}

\begin{figure}[t]
\begin{center}
\includegraphics[width=0.45\textwidth]{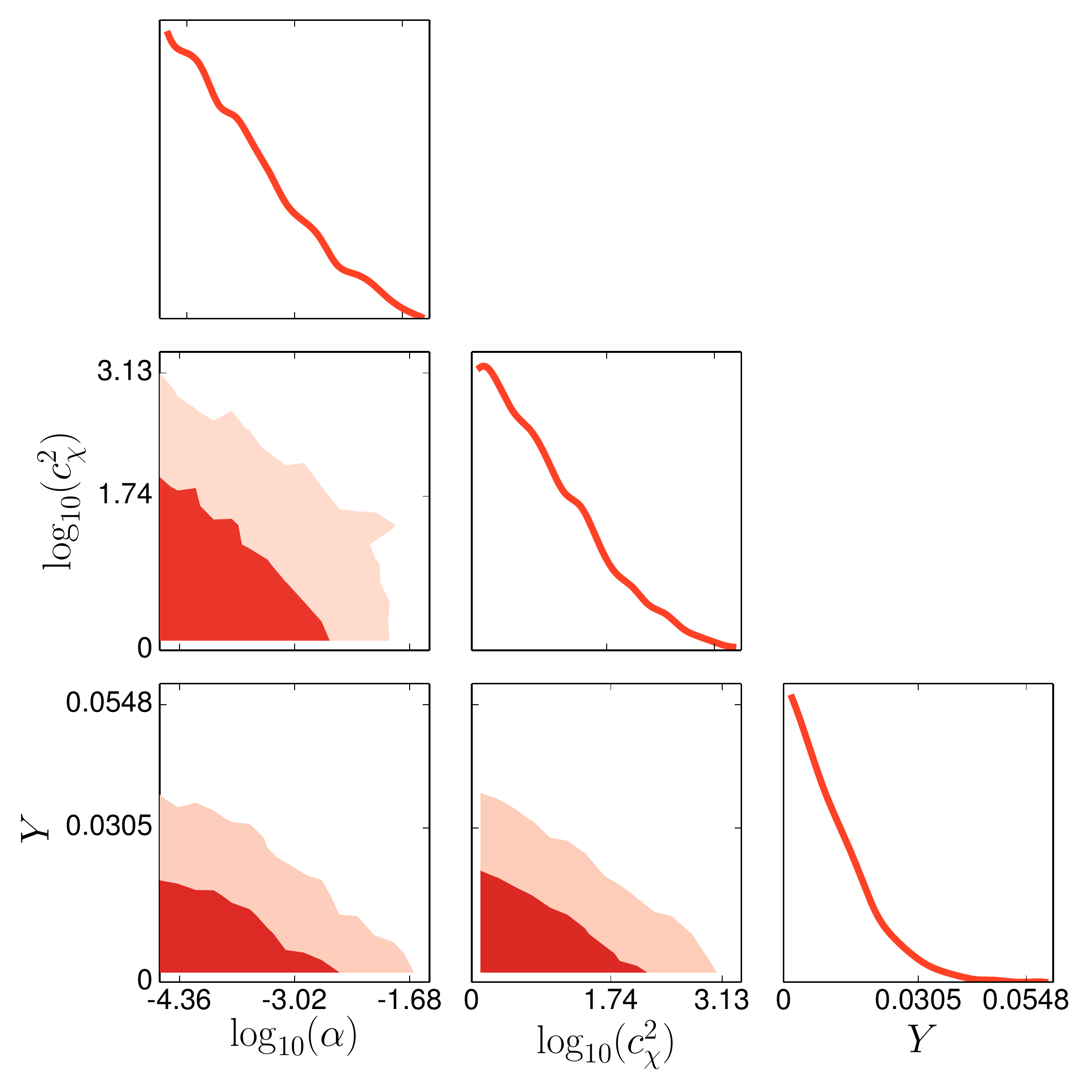}
\includegraphics[width=0.45\textwidth]{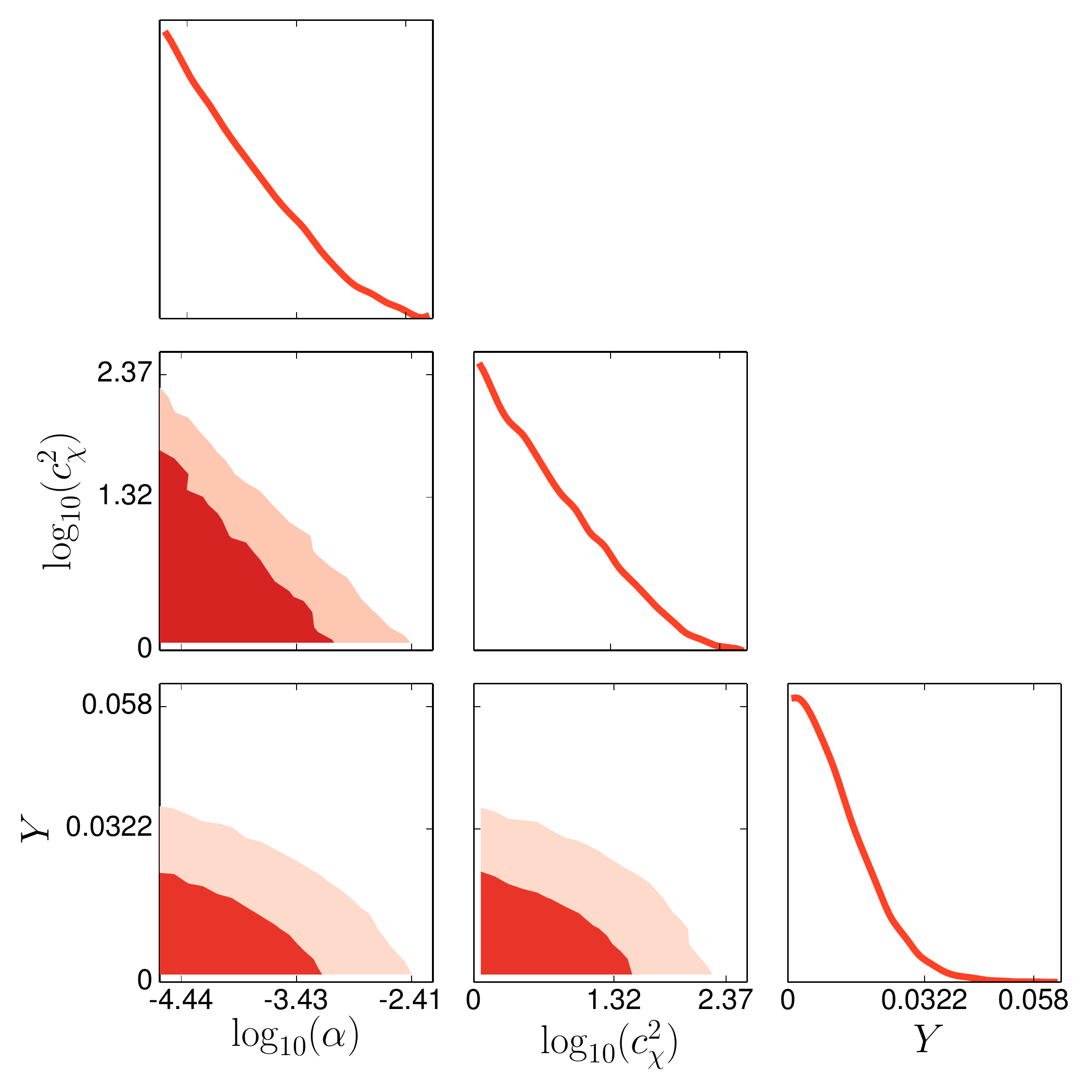}
\end{center}
\caption{\label{fig:deg2} Marginalized one-dimensional posterior
  distribution and two-dimensional probability contours (at the 68\%
  and 95\% CL) of the $\Lambda$LVDM
parameters for
Einstein-aether (upper panel) and khronometric (lower panel)
cases. Only the subspace of 
parameters responsible for Lorentz violation is shown.} 
\end{figure}

The effects discussed above are
constrained by the data on the CMB anisotropies and LPS. In this way
the cosmological observations can be used to put bounds on LV in dark
matter and gravity. For this purpose we use   
in this work the CMB data from the Planck 2013 release
\cite{Planck} combined with the  
galaxy power spectrum from the WiggleZ
redshift survey \cite{WiggleZ}. 
To compute the predictions for observables at various values of the
model 
parameters, we modify the  Boltzmann code {\sc Class}
\cite{Blas:2011rf}. 
The parameter space is explored using the Monte Carlo code {\sc Monte Python}
\cite{Audren:2012wb}. 
We consider both Einstein-aether and khronometric 
models focusing only on the scalar sector of perturbations. 
We separately study the cases $Y\neq 0$ (Lorentz violation in dark
matter) and $Y\equiv 0$ (dark matter is Lorentz invariant).

For non-zero $Y$ the fit includes nine cosmological parameters, which are
the usual six free parameters of the minimal flat $\Lambda$CDM model,
plus three parameters  
describing LV in gravity and dark matter, namely $\log_{10} \alpha$,
$\log_{10} c_{\chi}^2$ and $Y$. 
These combinations and the logscale
have been
chosen to improve the convergence of the Monte Carlo chains.
For the last two parameters, we impose a flat prior in the ranges
\bseq
\label{priors}
\be
0\leq\log_{10}c^2_\chi\;,~~~~
0<Y<1\;.
\ee
The
lower prior on $\log_{10} c_\chi^2$ follows from the Cherenkov bound
discussed in Sec.~\ref{sec:cosmo}.
We do not need an upper prior on $\log_{10} c_\chi^2$, 
as large values of $c_\chi^2$ are strongly
disfavored by the data, see Fig.~\ref{fig:deg2}. 
For given 
$\log_{10} \alpha$ and
$\log_{10} c_{\chi}^2$,
the parameters ($\beta$, $\lambda$) are fixed either by the condition 
(\ref{consaeth1}) in the Einstein-aether case or by (\ref{conskhron}) 
for the khronometric model. 
We have seen that the effects of LV in DM are screened 
at wavenumbers smaller 
than  
$k_{Y,0}\propto \sqrt{Y/(\a c_\chi^2)}$.
When $\a<0.5\cdot 10^{-5}$ (weak LV in gravity) one does not expect
to obtain bounds on LV in DM from CMB or LPS: any $Y\sim 1$ 
is allowed because the screening occurs over the whole   
range of scales relevant for these observables (we take 
$k_{max}=0.1\;h\cdot\mathrm{Mpc}^{-1}$ as the upper limit on the
region where the power spectrum can be considered as linear).
Thus, in order to obtain efficient constraints on $Y$, we impose a
flat prior in the range\footnote{We have found that the 
linear observables remain sensitive to the effects of LV in DM even 
for $k_{Y,0}$ somewhat above $k_{max}$, suggesting that some constraints
on $Y$ could be obtained down to the values $\a\sim 10^{-7}$. 
However, for large $k_{Y,0}$ the last term in brackets in 
eq.~(\ref{eq:chi}) leads to fast oscillations of the $\chi$-field and
the numerical 
computations become very demanding. Our decision  to
impose the stronger priors \eqref{prior2} represents a compromise
between the desire to explore the physically 
interesting portion of the parameter space and computational efficiency.}     
\be
\label{prior2}
-4.7< \log_{10} \a  < 0.3\;
\ee
\eseq
(which corresponds to $2\cdot 10^{-5}<\a < 2$).

\begin{figure}[h]
\begin{center}
\includegraphics[width=0.33\textwidth]{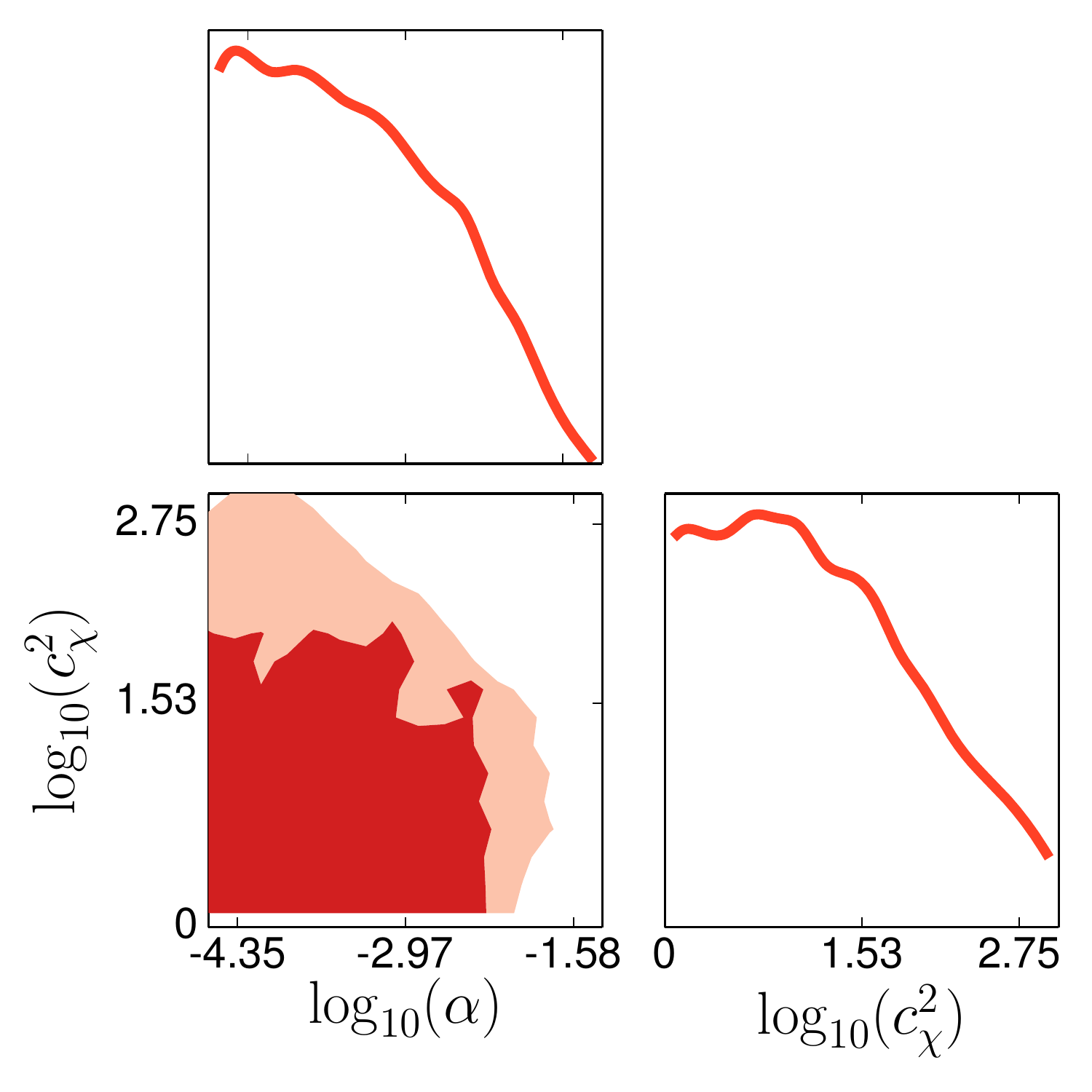}
\includegraphics[width=0.33\textwidth]{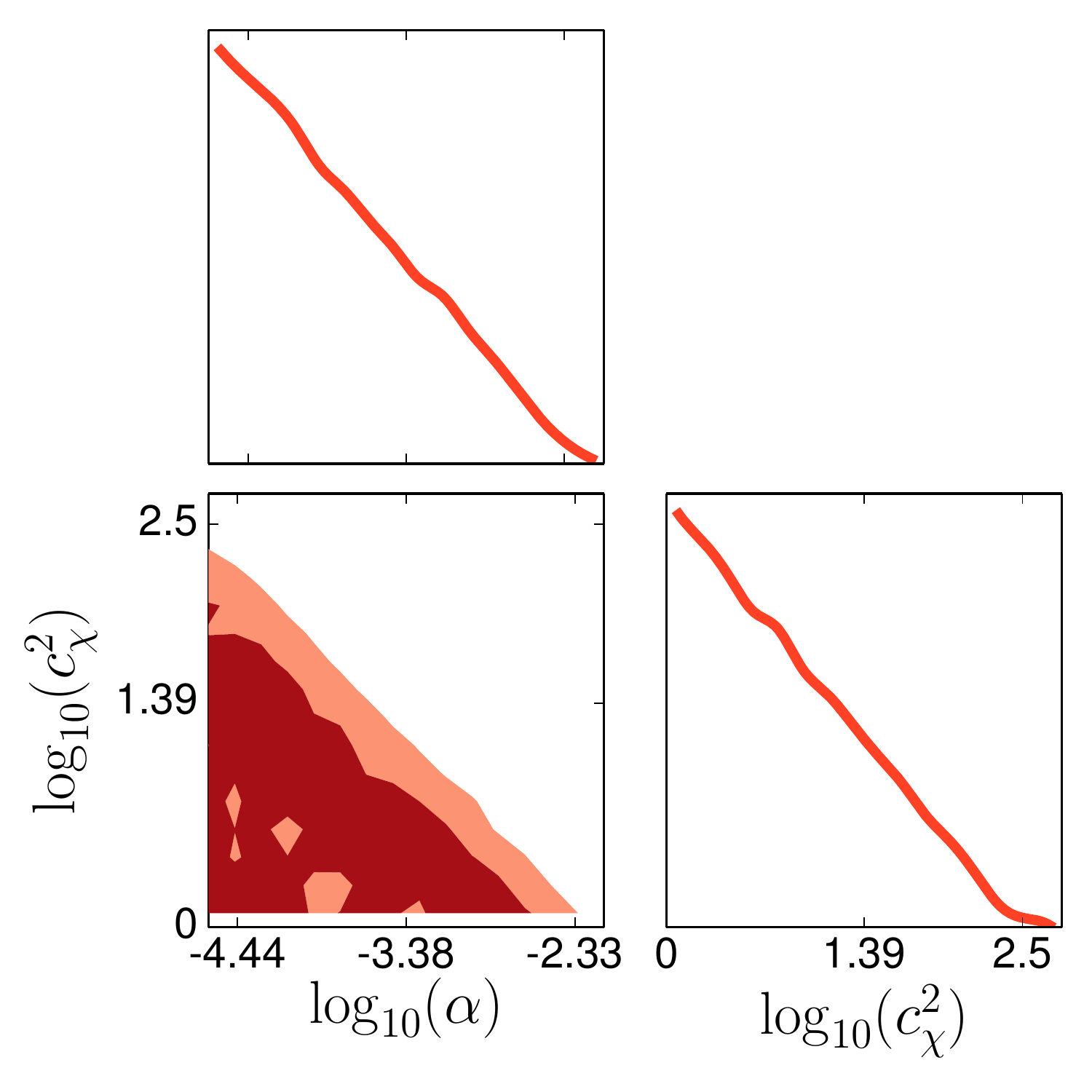}
\end{center}
\caption{\label{fig:deg3} Marginalized one-dimensional posterior
  distribution and two-dimensional probability contours (at the 68\%
  and 95\% CL) of the 
parameters for
Einstein-aether (upper panel) and khronometric (lower panel)
theories in the case of Lorentz invariant dark matter ($Y\equiv 0$). 
Only the subspace of 
parameters responsible for Lorentz violation is shown.} 
\end{figure}

Our results for the marginalized Bayesian minimum credible
intervals for the parameters of the fit are presented in the first two
lines of Table~\ref{table}.   
The three LV parameters are found to be very weakly correlated with
the usual 
$\Lambda$CDM parameters.
The one-dimensional and two-dimensional posterior parameter distributions 
for LV parameters are displayed in Fig.~\ref{fig:deg2}. We see that
the posterior distributions for $\log_{10}\a$,  $\log_{10}c_\chi^2$
and $Y$ are peaked at the lower ends of the scanned parameter space,
implying that the data bring no evidence for LV. In particular, the
parameter $Y$ governing LV in the DM sector is constrained to be less
than $Y<0.03$ in both models. We stress, though, 
that this bound should be taken in conjunction with the priors
(\ref{priors}). As we already explained, the linear observables become
insensitive to LV in DM for very small values of   
$\alpha$. The data on non-linear structure 
can presumably  help to find constraints in this regime. However, this
requires an analysis of the non-linear dynamics of the model, which is
beyond the scope of the present paper.

We observe a triangular shape for the allowed region in the 
$(\log_{10} \a, \log_{10} c_\chi^2)$ plane. One concludes
that the data essentially constrain the
combination 
$\alpha c_\chi^2$ entering $k_{Y,0}$
by disfavoring models where the enhanced growth of structure due to
LV in DM takes place on linear scales. 
This implies that the introduction of a non-zero
parameter $Y$ in the fit biases the combination $\a c_\chi^2$
towards smaller values.
Therefore, the bounds on $\a$ and $c_\chi^2$ from the
first two lines in Table~\ref{table} do not apply to the case when LV
is confined to the gravitational sector only and DM is exactly Lorentz
invariant ($Y\equiv 0$ from the start). To obtain the constraints in
this case we ran separate series of simulations setting $Y=0$ in all
dynamical equations. Otherwise we followed the same procedure as
before. We varied $\log_{10}\a$ and $\log_{10}c_\chi^2$ in addition to
the six standard  
$\Lambda$CDM parameters, with the priors \eqref{priors}.
The resulting
posterior distributions for the parameters 
$(\log_{10}\a,\log_{10}c_\chi^2)$ are shown in Fig.~\ref{fig:deg3} and
the marginalized credible intervals are listed in the last two lines of
Table~\ref{table}. One observes that the new bounds are indeed weaker
than in the case of LV DM.

For the khronometric model (Fig.~\ref{fig:deg3}, lower panel) the
isoprobability contours have the same shape as in the case of
$\Lambda$LVDM implying that the 
constrained 
combination is again $\a
c_\chi^2$. This is consistent with the expectation that the main
effects in this case are related to the difference between $G_N$ and
$G_{cos}$ (type {\em (i)} according to the classification of
Sec.~\ref{sec:effects}). Imposing the PN constraint (\ref{conskhron})
one obtains,
\be
\frac{G_N}{G_{cos}}-1=\frac{3}{2}\a c_\chi^2\;,
\ee
where we have used (\ref{selfgravity}) and the first formula in
(\ref{ckY}). This is the combination mostly constrained by the data.
Note that the bound on $\a$ obtained in this paper is 50 times better
than a bound derived in Ref.~\cite{Audren:2013dwa} for a similar
model (the bound in \cite{Audren:2013dwa} is formulated in
  terms of $\b=\a/2$ and reads $\b<0.05$). 
This is a consequence of the reduction of the parameter space
after imposing
the Cherenkov radiation constraint
$c_\chi\geq 1$. Another source of improvement
is the use of more recent CMB data (Planck 2013 vs. WMAP 7 and SPT in
\cite{Audren:2013dwa}).

In the Einstein-aether case (Fig.~\ref{fig:deg3}, upper panel) 
the bounds on $\a$
and $c_\chi^2$ are significantly weaker. This is explained by the
observation made in Sec.~\ref{sec:effects} that, once the relevant PN
constraint (\ref{consaeth1}) is imposed, the effects of enhanced
gravity disappear and we are left with the effects of the type {\em (ii)}
related to the shear sourced by the aether. These effects are
proportional to 
$\beta=\alpha(1+3 c_{\chi}^2)/2$ and are rather suppressed, which
translates into mild bounds on $\a$ and $c_\chi^2$. Note, however,
that the bound on $\a$ is still at the per cent level, which is
comparable with the bounds from binary pulsars \cite{Yagi:2013qpa}.

\section{Conclusions}
\label{sec:6}

In this paper we have used the Planck 2013 and WiggleZ 2012 data to
derive constraints on the deviations from Lorentz invariance in
gravity and dark matter. We considered the scenario where local
Lorentz invariance is broken down to spatial rotations preserving a time-like
direction. This pattern of symmetry breaking is described at low
energies by the Einstein-aether or khronometric model. The latter case
represents the infrared limit of Ho\v rava gravity. We allowed
for a possible LV coupling between aether (khronon) and dark matter
keeping Lorentz invariance in the sectors of the Standard Model and
dark energy (accounted for by a cosmological constant). 

While for the background cosmological evolution the resulting
$\Lambda$LVDM model is almost equivalent to $\Lambda$CDM, the
difference is substantial at the
level of perturbations. We studied the impact of these differences on
the CMB and LPS using the modified 
Boltzmann code {\sc Class} \cite{Blas:2011rf},
and explored the parameter space of $\Lambda$LVDM with the parameter
inference code 
{\sc Monte Python} \cite{Audren:2012wb}. We performed four series of
Monte Carlo simulations for the Einstein-aether and khronometric cases
with and without LV in dark matter. In the analysis we
imposed as priors the constraints from local gravitational
measurements in the Solar System and the astrophysical bound following
from the absence of vacuum Cherenkov losses by ultra-high energy
cosmic rays. As a result we have obtained the most stringent
constraints to date on the departures from Lorentz invariance in
cosmology. Our limits on the parameters of LV in the gravity sector
are competitive with those coming from the slow-down of binary pulsars
\cite{Yagi:2013qpa}. The dimensionless parameter $Y$ characterizing LV in
dark matter
has
been constrained to be less than $0.03$. This parameter 
has the meaning of the difference between the
maximal velocity\footnote{Here, we are referring to the maximal velocity that DM particles could reach in principle in the ultra-relativistic limit, and not to actual DM velocities in the Universe during matter domination.} of dark matter particles and the speed of light.
Our analysis provides the 
first direct constraint on this quantity.

It should be emphasized, however, that the latter constraint applies only 
under the priors (\ref{priors}) imposed to keep  
the effects of LV in dark matter unscreened in the linear regime.
When the screening happens,
CMB and LPS 
get insensitive to the parameter $Y$. It will be interesting to
investigate if in this regime LV dark matter can still be tested
through the properties of the non-linear structures, such as e.g. the halo
mass function. 

In our analysis we focused on the scalar sector of perturbations. The
presence of vector excitations in the Einstein-aether case can
affect  
the CMB polarization, including the B-mode
\cite{ArmendarizPicon:2010rs,Nakashima:2011fu}. We leave the study of
the corresponding constraints for the future.\\
\\


\textit{Acknowledgments.} We are grateful to Grigory Rubtsov and Oleg
Ruchayskiy for
useful discussions. 
The numerical simulations were partially performed 
at the cluster of the Theoretical Division of INR RAS.
This work was supported in
part by the Swiss National Science Foundation (B.A., M.I., J.L., S.S.).
M.I. also acknowledges support from the
Grants of the President of Russian Federation 
MK-1754.2013.2, NS-2835.2014.2, and the RFBR grants 
14-02-31435 and 14-02-00894.

\end{document}